\documentclass[aps,floats]{revtex4}
\usepackage{amsmath,amssymb}
\usepackage{graphicx,epsfig}

\begin{document}
\bibliographystyle {plain}

\def\oppropto{\mathop{\propto}} 
\def\opsimeq{\mathop{\simeq}}
\def\opoverderline{\mathop{\overline}}
\def\operarrow{\mathop{\longrightarrow}}
\def\opsim{\mathop{\sim}}

\def\fig#1#2{\includegraphics[height=#1]{#2}}
\def\figx#1#2{\includegraphics[width=#1]{#2}}


\title{  Fractal dimension of  spin glasses interfaces in dimensions $d=2$ and $d=3$ \\
via strong disorder renormalization at zero temperature
  } 


 \author{ C\'ecile Monthus }
  \affiliation{Institut de Physique Th\'{e}orique,\\
  CNRS and CEA Saclay \\
 91191 Gif-sur-Yvette, France}

\begin{abstract}
For Gaussian Spin Glasses in low dimensions, we introduce a simple Strong Disorder renormalization procedure at zero temperature. In each disordered sample, the difference between the ground states associated to Periodic and Anti-Periodic boundary conditions defines a system-size Domain Wall. The numerical study in dimensions $d=2$ (up to sizes $2048^2$) and $d=3$ (up to sizes $128^3$) yields fractal Domain Walls of dimensions $d_s(d=2) \simeq 1.27$ and $d_s(d=3)  \simeq 2.55$ respectively.

\end{abstract}

\maketitle

\section{ Introduction}

Within the droplet scaling theory of classical spin-glasses in finite dimensions $d$\cite{mcmillan,bray_moore,fisher_huse}, the two universal critical exponents that characterize the zero-temperature fixed point can be defined by considering, in each disordered sample ${\cal J}$ of volume $L^d$, 
the two ground states associated to two different boundary conditions, 
 for instance Periodic (P) and Anti-Periodic (AP).
The difference between the two ground states
 defines a system-size Domain-Wall :

(i) the scaling of its energy $E^{DW}$ with respect to the linear size $L$
defines the droplet or stiffness exponent $\theta$
\begin{eqnarray}
E^{DW}_{\cal J} \equiv  E^{GS(AP)}_{\cal J}-E^{GS(P)}_{\cal J}  = L^{\theta} u
\label{edw}
\end{eqnarray}
where $u$ is an $O(1)$ random variable of zero mean.
The numerical values measured in dimensions $d=2$ and $d=3$ read
(see \cite{boettcher} and references therein)
\begin{eqnarray}
\theta(d=2) && \simeq - 0.28
\nonumber \\
\theta(d=3) && \simeq 0.24
\label{thetahypercubic}
\end{eqnarray}

(ii) the scaling of its surface $\Sigma^{DW} $ (number of bonds belonging to the domain wall) defines the fractal dimension $d_s$ of the Domain-Wall
\begin{eqnarray}
\Sigma^{DW}_{\cal J}   \propto L^{d_s} v
\label{sdw}
\end{eqnarray}
where $v$ is an $O(1)$ positive random variable.
The numerical values measured in dimensions $d=2$ 
\cite{BrayMoore_chaotic,middleton,hartmann,melchert,sle_amoruso,sle_bernard,roma} 
and $d=3$ \cite{palassini,katzgraber} read
\begin{eqnarray}
d_s(d=2) && \simeq 1.28
\nonumber \\
d_s(d=3) && \simeq 2.58
\label{dshypercubic}
\end{eqnarray}
Moreover in dimension $d=2$, the Domain-Wall of fractal dimension $d_s \simeq 1.28$
has been characterized as an SLE process \cite{sle_amoruso,sle_bernard}.
The fractal dimension $d_s$ plays in particular a major role in the chaos properties of spin-glasses with respect to temperature and to disorder perturbations \cite{bray_moore,fisher_huse}, as well as in the dynamics \cite{us_conjecture}.

From the point of view of real-space renormalization, 
spin-glasses have been mostly studied within
 the Migdal-Kadanoff approximation 
\cite{southern_young,young_St,mckay,BM_MK,Gardnersg,banavar,bokil,muriel,thill,ritortMK,boettcher_mk,us_tail,jorg,us_sgferro} where the hypercubic lattice is effectively replaced by a hierarchical
fractal lattice whose structure is exactly renormalizable
by construction
\cite{MKRG,berker,hierarchical} : this approximation  
reproduces very well
the values of the droplet exponent in dimensions $d=2$ and $d=3$
 (Eq \ref{thetahypercubic}), but not the surface dimension which is 
fixed to the trivial value
\begin{eqnarray}
d_s^{MK} =d-1
\label{dsMK}
\end{eqnarray}
If one insists on keeping the
hypercubic lattice in dimension $d>1$, the precise definition of an
appropriate renormalization procedure has remained very difficult.
The reason can be understood by considering a standard Block Renormalization
using the maximal Block size $b=\frac{L}{2}$ :
 if one decomposes the volume $L^d$ into $2^d$ blocks
of volume $\left(\frac{L}{2}\right)^d$ and compute the ground states
in each block, the residual coupling between two blocks is the sum
over $\left(\frac{L}{2}\right)^{d-1}$ initial couplings of random signs,
 leading to the too high value
\begin{eqnarray}
\theta^{Block} && = \frac{d-1}{2}
\label{thetablock}
\end{eqnarray}
This poor value comes from the facts
 that the Domain-Walls have been assumed to be of dimension 
\begin{eqnarray}
d_s^{Block} && =d-1
\label{dsblock}
\end{eqnarray}
and have been fixed at the middle of the sample independently of the disorder realization. To obtain better results, it is thus necessary
to build correlated clusters with boundaries adapted to each disorder realization.
In the present paper, we thus introduce and study numerically
some simple Strong Disorder real-space renormalization in dimensions
 $d=2$ and $d=3$, and obtain that it is able to reproduce very
 well the fractal dimensions $d_s$ quoted in Eq. \ref{dshypercubic}.

The paper is organized as follows.
The principle of the Strong Disorder real-space renormalization
at zero temperature is described in section \ref{sec_rg}.
The numerical application in dimension $d=2$ and $d=3$ are presented in sections
\ref{sec_rg2d} and \ref{sec_rg3d} respectively.
Our conclusions are summarized in section \ref{sec_conclusion}.

\section{ Strong Disorder Renormalization at zero temperature }

\label{sec_rg}

For a finite-dimensional spin-glass of Hamiltonian
\begin{eqnarray}
H && = - \sum_{(i,j)} J_{ij} S_i S_j
\label{hsg}
\end{eqnarray}
where $S_i = \pm 1$ are classical spins,
and where $J_{ij}$ are the random Gaussian couplings of zero mean,
 we would like to construct the ground state via 
some simple strong disorder real-space renormalization procedure. 

\subsection{ Analysis of the local fields }

For each spin $S_i$, we thus consider its local field
\begin{eqnarray}
h^{loc}_i && = \sum_{j} J_{ij}  S_j
\label{hloci}
\end{eqnarray}
and compute 
its largest coupling in absolute value, corresponding to some index $j_{max}(i)$
\begin{eqnarray}
 \max_{j} ( \vert J_{ij}  \vert  ) \equiv \vert J_{i,j_{max}(i)}  \vert
\label{omegai}
\end{eqnarray}
We ask whether the local field
\begin{eqnarray}
h^{loc}_i && =  J_{i,j_{max}(i)}   S_{j_{max}(i)}+ \sum_{j \ne j_{max}(i)}  J_{ij} S_j
\label{hloci2}
\end{eqnarray}
 can be dominated by the first term whatever the values taken by the spins $S_j$
of the second term.

\subsection{ Comparison with the worst case }

The 'worst case' 
is of course when all the spins $S_j$ of the second term in Eq. \ref{hloci2}
are such that
$( J_{ij} S_j)$ all have the same sign, so that their contribution to the local field is maximal. It is thus convenient to introduce the difference
\begin{eqnarray}
\Delta_i && \equiv \vert J_{i,j_{max}(i)}  \vert - \sum_{j \ne j_{max}(i)}  \vert J_{ij}  \vert
\label{deltai}
\end{eqnarray}

If $\Delta_{i_0}>0$, the sign of the local field $h^{loc}_{i_0}$ 
will be determined by the sign of the first term $J_{i_0j_{max}(i_0)}  S_{j_{max}(i_0)} $
for all values taken by the other spins $S_j$ with $j \ne j_{max}(i_0) $
\begin{eqnarray}
{\rm sgn} ( h^{loc}_{i_0} ) && =S_{j_{max}({i_0})} {\rm sgn} ( J_{{i_0},j_{max}({i_0})}  )
\label{hlocisgn}
\end{eqnarray}
Then the spin $S_{i_0}$ can be eliminated via
\begin{eqnarray}
S_{{i_0}} =  S_{j_{max}({i_0})} {\rm sgn} (J_{{i_0} j_{max}({i_0})})
\label{elimsi0}
\end{eqnarray}
so that the Hamiltonian of Eq. \ref{hsg} becomes
\begin{eqnarray}
H && = - \vert J_{{i_0} j_{max}({i_0})} \vert - \sum_{(i,j)\ne i_0} J_{ij}^R S_i S_j
\label{hsgdeci}
\end{eqnarray}
where the renormalized couplings concerning the spin $S_{j_{max}(i_0)}$ read
\begin{eqnarray}
J^R_{j_{max}(i_0),j} = J_{j_{max}(i_0),j}+ J_{i_0,j}  {\rm sgn} (J_{i_0 j_{max}(i_0)})
\label{jr}
\end{eqnarray}

For a site $i_0$ having coordinence $z=2$ (i.e.
only two neighbors), the difference $\Delta_{i_0}$
is always positive, so that the renormalization above is exact :
this is the case in particular in dimension $d=1$ and in the Migdal-Kadanoff
approximation in dimension $d>1$ (see sections 4.2 and 4.3
of Ref \cite{rgLRSG} for more details).

\subsection{ Comparison with the typical case }

However ``the worst is not always true'' :
indeed in a frustrated spin-glass, the worst case discussed above
where all the spins $S_j$ are such that $( J_{ij} S_j)$ have all the same sign,
 is expected to be rather atypical. In the absence of other informations, 
it is much more natural to compare with a sum of random 
terms of absolute values $J_{ij}$ and of random signs, i.e. to replace the difference $\Delta_i$ of Eq. \ref{deltai}
by
\begin{eqnarray}
\Omega_i && \equiv \vert J_{i,j_{max}(i)}  \vert - \sqrt{ \sum_{j \ne j_{max}(i)}  \vert J_{ij}  \vert^2  }
\label{omegaii}
\end{eqnarray}
Note that for the case of coordinence $z=2$, $\Omega_i$ actually
coincides with $\Delta_i$,
so that the exactness discussed above is the same. 
But for coordinence $z>2$, we expect that $\Omega_i$ is a better indicator of
the relative dominance of the maximal coupling for the different spins.
We have thus chosen to introduce the Strong Disorder Renormalization
procedure based on the variable $\Omega_i$ as we now describe.

\subsection{ Formulation of Strong Disorder Renormalization procedure }

At each step, the spin-glass Hamiltonian of the form of Eq. \ref{hsg}
contains $N$ remaining spins. Each spin $S_i$ is characterized
  by the variable $\Omega_i$ of Eq. \ref{omegaii} computed from the couplings $J_{ij}$ connected to $S_i$.

 The iterative renormalization procedure is defined by
the following elementary decimation step  :

 Find the spin $i_0$ with the maximal $\Omega_i$
\begin{eqnarray}
 \Omega_{i_0} \equiv \max_{i} (  \Omega_{i}    ) 
\label{omegaimax}
\end{eqnarray}
 The elimination of the spin $S_{i_0}$ via the rule of Eq. \ref{elimsi0}
yields that all its couplings $J_{i_0,j} $ with $j \ne j_{max}(i_0)$ 
are transferred to the spin $S_{j_{max}(i_0)}$ via the renormalization rule
of Eq. \ref{jr}
\begin{eqnarray}
J^R_{j_{max}(i_0),j} = J_{j_{max}(i_0),j}+ J_{i_0,j}  {\rm sgn} (J_{i_0 j_{max}(i_0)})
\label{rulejr}
\end{eqnarray}

The procedure ends when only a single spin $S_{last}$ is left :
the two values $S_{last}=\pm 1$ label the two ground states related by a global flip of all the spins.
From the choice $S_{last}=+1$, we may reconstruct all the values of the
decimated spins via the rule of Eq. \ref{elimsi0}.

In the sections \ref{sec_rg2d} and \ref{sec_rg3d}, we study numerically this renormalization procedure
in dimensions $d=2$ and $d=3$ to see whether the corresponding exponents $\theta$
and $d_s$ are closer to the numerical values of Eqs \ref{thetahypercubic}
and \ref{dshypercubic} than the block values of Eqs \ref{thetablock} and
\ref{dsblock}. But let us first mention similarities and differences with previous works.

\subsection{ Differences with the 'greedy' procedure for classical spin-glasses }

The simplest 'greedy' procedure introduced for classical spin-glasses
consists in satisfying the bonds in the order of the absolute values of the couplings $\vert J_{i,j} \vert $ unless a closed loop appears \cite{cieplak,NSgreedy,jackson}. So the two differences with the present procedure is that

(i) here we decimate spins according to the biggest
 variable $\Omega_i$ of Eq. \ref{omegaii}, instead 
of decimating 'bonds' according to the biggest $\vert J_{i,j} \vert $

(ii) here the couplings are renormalized according to Eq. \ref{rulejr},
whereas in the greedy procedure of Refs \cite{cieplak,NSgreedy,jackson},
no renormalization is mentioned.

In dimensions $d=2$ and $d=3$, the corresponding fractal dimensions of 
the Domain-Wall have been numerically measured to be $d^{greedy}_f(d=2) \simeq 1.2 \pm 0.02$ 
and $d^{greedy}_f(d=3) \simeq 2.5 \pm 0.05$ \cite{cieplak}
(i.e. somewhat slightly lower than Eq. \ref{dshypercubic} at least in dimension $d=2$).

In the context of the one-dimensional spin-glasses with power-law couplings
$J(r) \propto 1/r^{\sigma}$,
we have studied recently a related strong disorder renormalization procedure
able to reproduce the correct droplet exponent \cite{rgLRSG},
using the strong hierarchy of initial couplings with the distance.

\subsection{ Differences with Strong Disorder RG for quantum spin-glasses }

As a final remark, we should also stress the difference with the 
strong disorder renormalization method
 (see \cite{review_strong} for a review) that has been developed for
disordered {\it  quantum } spin models either
 in $d=1$ \cite{fisher} or in $d=2,3,4$ 
 \cite{motrunich,fisherreview,lin,karevski,lin07,yu,kovacsstrip,kovacs2d,kovacs3d,kovacsentropy,kovacsreview}.
In these quantum spin models, the idea is to decimate the strongest coupling $J_{max}$ remaining in the whole system :  the renormalized couplings obtained via second order perturbation theory of quantum mechanics
are obtained as ratios of couplings and are thus
typically much weaker than the decimated coupling $J_{max}$, so that the procedure is consistent and even asymptotically exact at the critical point where the typical renormalized couplings {\it decays }  as $J_L^{typ} \propto e^{- L^{\psi}}$. 
This is thus completely different from the problem of {\it classical } 
spin-glasses at zero temperature considered in the present paper, 
where the droplet scaling is a power-law  $J_L \propto L^{\theta}$,
and where the renormalization rule is of the form of Eq. \ref{rulejr}, so that it is not easy to know in advance whether the procedure will be consistent or not. 
In the following sections, we thus present numerical results.

\section{ Application to the Gaussian spin-glass in dimension $d=2$ }

\label{sec_rg2d}

\subsection{ Numerical procedure }

\label{secnum2d}

For each disordered sample defined on a square lattice $L \times L$
\begin{eqnarray}
H_{2d} = -\sum_{x=1}^L \sum_{y=1}^L 
S_{(x,y)} \left[ J^{P}_x(x,y)S_{(x+1,y)} + J^{P}_y(x,y)S_{(x,y+1)}  \right]
\label{hsg2d}
\end{eqnarray}
with periodic boundary conditions in the two directions 
$S_{(L+1,y)} \equiv S_{(1,y)}$ and $S_{(x,L+1)} \equiv S_{(x,1)}$ 
we apply the renormalization procedure to construct the ground-state configuration
$\{S_i^{GS(P)}\}$. 
As usual, the Anti-Periodic Boundary conditions in the direction $x$
($S_{(L+1,y)} = - S_{(1,y)}$) can be equivalently studied by  
 changing the signs of the horizontal couplings in the column $x=1$
\begin{eqnarray}
 J_x^{AP}(x=1,y) = - J_x^{P}(x=1,y)
\label{ruleAP}
\end{eqnarray}
The renormalization procedure is again applied 
to construct the ground-state configuration $\{S_i^{GS(AP)}\}$. 
The number of bonds having a different satisfaction 
 between the two ground states
\begin{eqnarray}
S_i^{GS(AP)} S_j^{GS(AP)} {\rm sgn}(J^{AP}_{ij}) \ne S_i^{GS(P)} S_j^{GS(P)} {\rm sgn}(J^{P}_{ij})
\label{diffDW}
\end{eqnarray}
corresponds to the surface $\Sigma^{DW}_{\cal J}$ of the Domain-Wall (Eq. \ref{sdw}),
whereas its energy $E^{DW}_{\cal J}$ corresponds to the difference between the
two ground states energies of Eq. \ref{edw}.

\subsection{ Results in a given sample  }

\begin{figure}[htbp]
 \includegraphics[height=6cm]{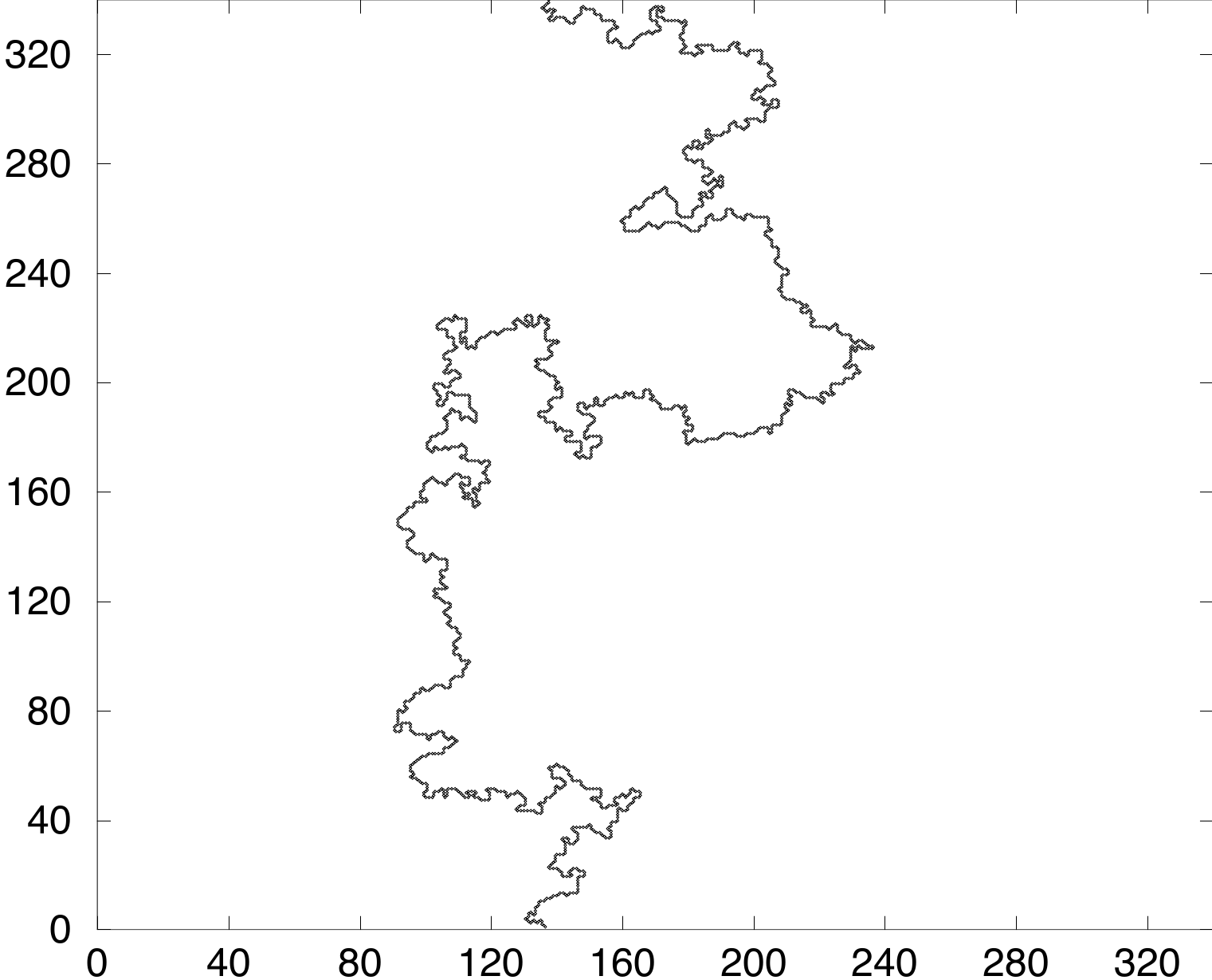}
\hspace{1cm}
 \includegraphics[height=6cm]{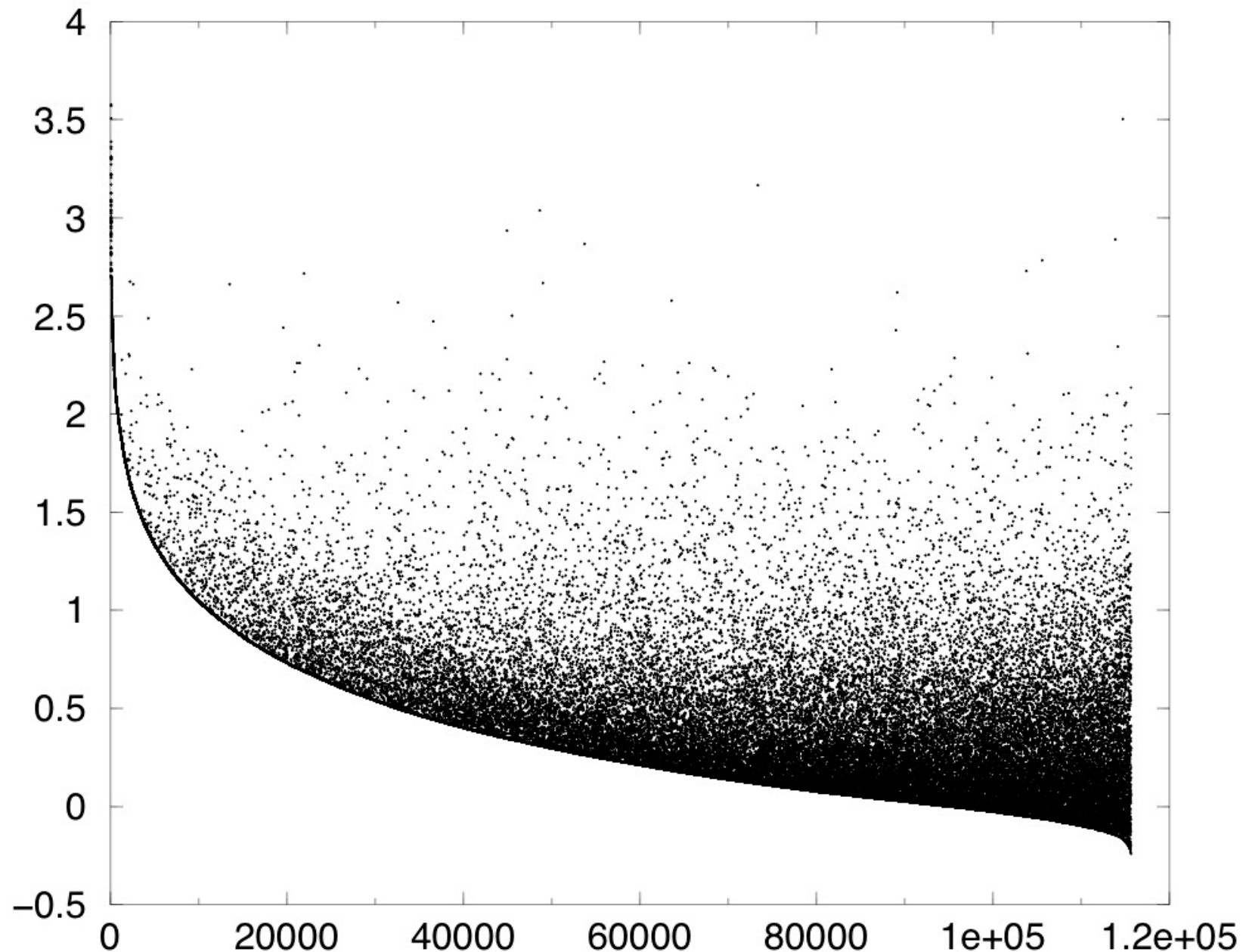}
\caption{ Strong Disorder Renormalization procedure in a two dimensional sample 
 of size $340 \times 340$ : \\
(a) Domain-Wall between the Periodic and the Anti-Periodic Boundary conditions. \\
(b) RG parameter $\Omega$ of the decimated spin as a function of the RG step
(the RG step corresponds to 
the number of spins that have already been decimated).
  }
\label{figinter2d}
\end{figure}

As an example, we show on Fig. \ref{figinter2d} (a)
 the interface obtained in a given sample by the procedure described above.

On Fig. \ref{figinter2d} (b), we also show the corresponding values of the RG parameter $\Omega$ of Eq. \ref{omegaii} as a function of the RG step.

\subsection{ Statistics over disordered samples }

\begin{figure}[htbp]
 \includegraphics[height=6cm]{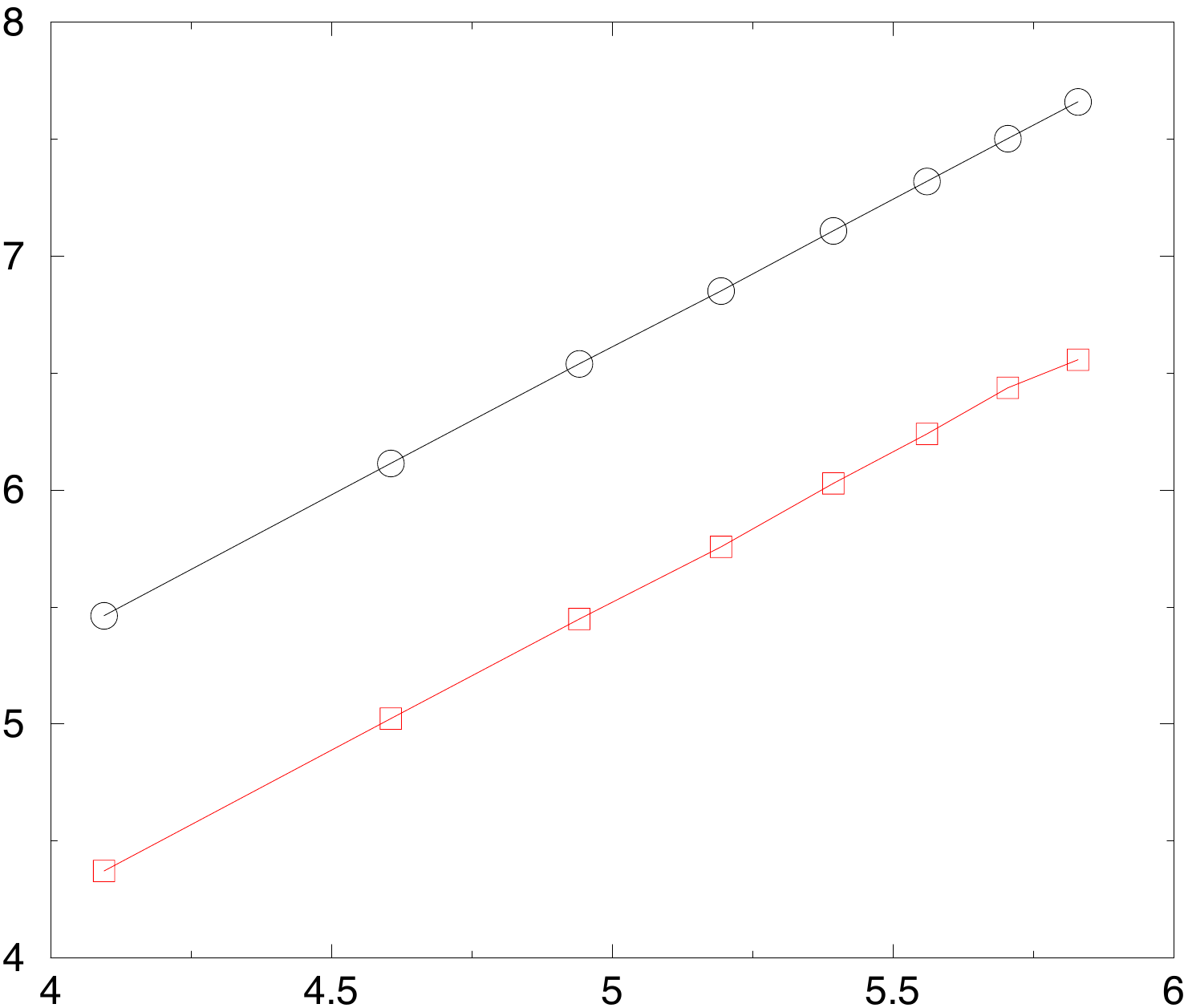}
\caption{ Log-log plot of the average value $ \overline{\Sigma^{DW} } $ (circles)
and of the width $\sqrt{ \overline{ (\Sigma^{DW})^2 }-(\overline{\Sigma^{DW} })^2 } $
(squares) of the length of the Domain-Wall as a function of the size $60 \leq L \leq 340$ of samples : the two slopes correspond to the fractal dimension $d_s \simeq 1.27$.  }
\label{figds2d}
\end{figure}

The application to $n_s(L)$ independent disordered samples
of various sizes $L$ with
\begin{eqnarray}
 L && = 60, 100, 140, 180, 220, 260, 300, 340 \nonumber \\
n_s(L) && = 4 \times 10^6,  4 \times 10^5 , 8 \times 10^4 , 25 \times 10^3 ,
 10^4, 4 \times 10^3, 2\times 10^3, 12 \times 10^2
\label{nume2d}
\end{eqnarray}
yields that the average value 
and the width of the length $\Sigma^{DW} $ of the Domain-Wall
have the same scaling (see Fig. \ref{figds2d})
\begin{eqnarray}
 \overline{\Sigma^{DW} } && \propto L^{d_s}  \nonumber \\
\sqrt{ \overline{ (\Sigma^{DW})^2 }-(\overline{\Sigma^{DW} })^2 } && \propto L^{d_s}
\label{sigma2d}
\end{eqnarray}
so that the Domain Wall is a fractal curve of dimension
\begin{eqnarray}
d_s \simeq 1.27
\label{numeds2d}
\end{eqnarray}
in agreement with the value quoted in Eq. \ref{dshypercubic} measured via exact numerical methods \cite{BrayMoore_chaotic,middleton,hartmann,melchert,sle_amoruso,sle_bernard,roma}.

However the droplet exponent that we measure from the width of the distribution
of the Domain-Wall energy $ E^{DW}$ (Eq. \ref{edw})
\begin{eqnarray}
\sqrt{ \overline{ (E^{DW})^2 } } \propto L^{\theta} 
\label{theta2dmes}
\end{eqnarray}
nearly vanishes $\theta \simeq 0$, i.e. it is far from the correct negative
value $\theta (d=2) \simeq -0.28$ quoted in Eq. \ref{thetahypercubic}, even if it is not as bad
as the positive Block value $\theta^{Block}(d=2)  = \frac{1}{2} $ of Eq. \ref{thetablock}.

\subsection{ Box-variant of the Strong Disorder Renormalization procedure } 

\label{secboxvariant2d}

\begin{figure}[htbp]
 \includegraphics[height=6cm]{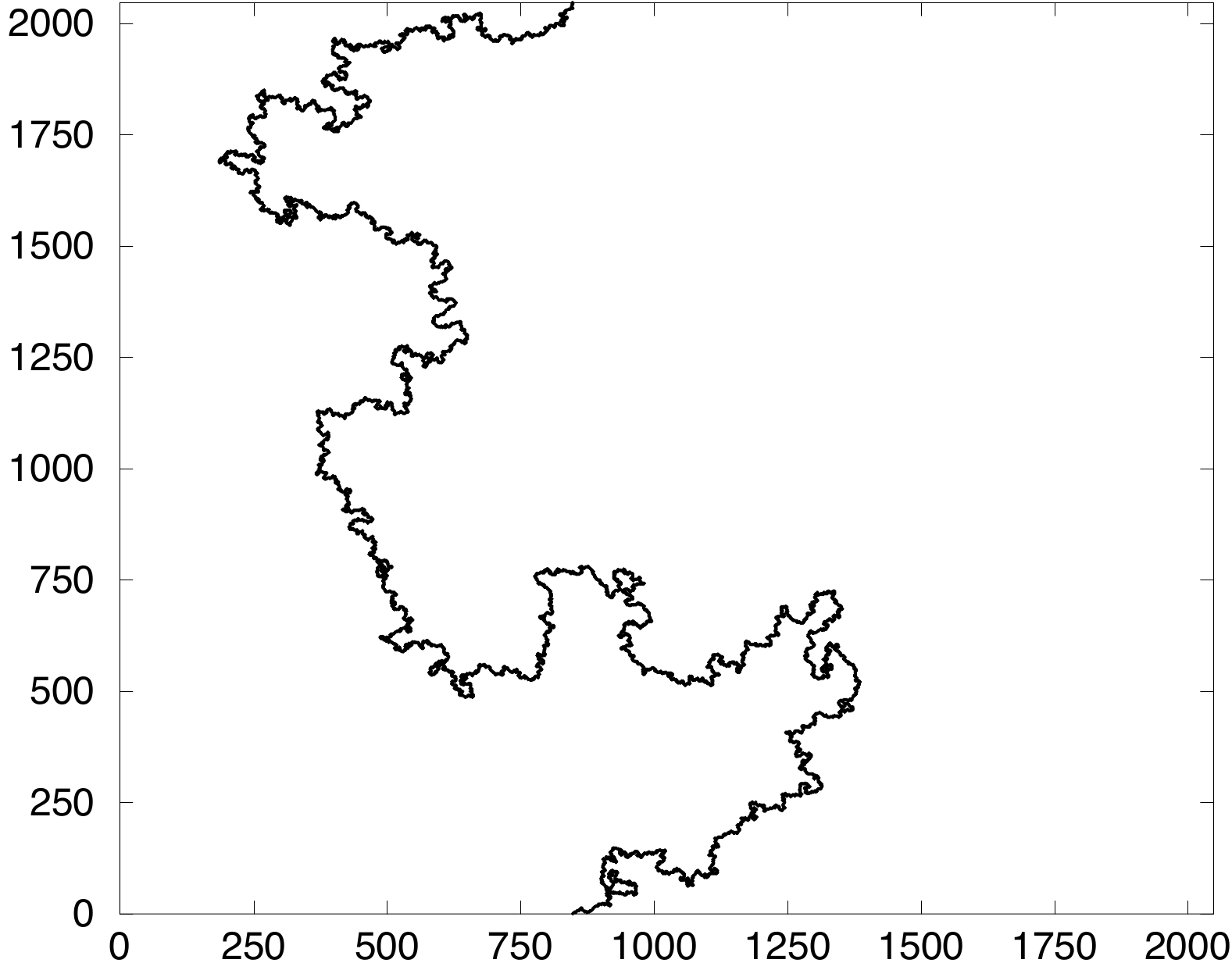}
\hspace{1cm}
 \includegraphics[height=6cm]{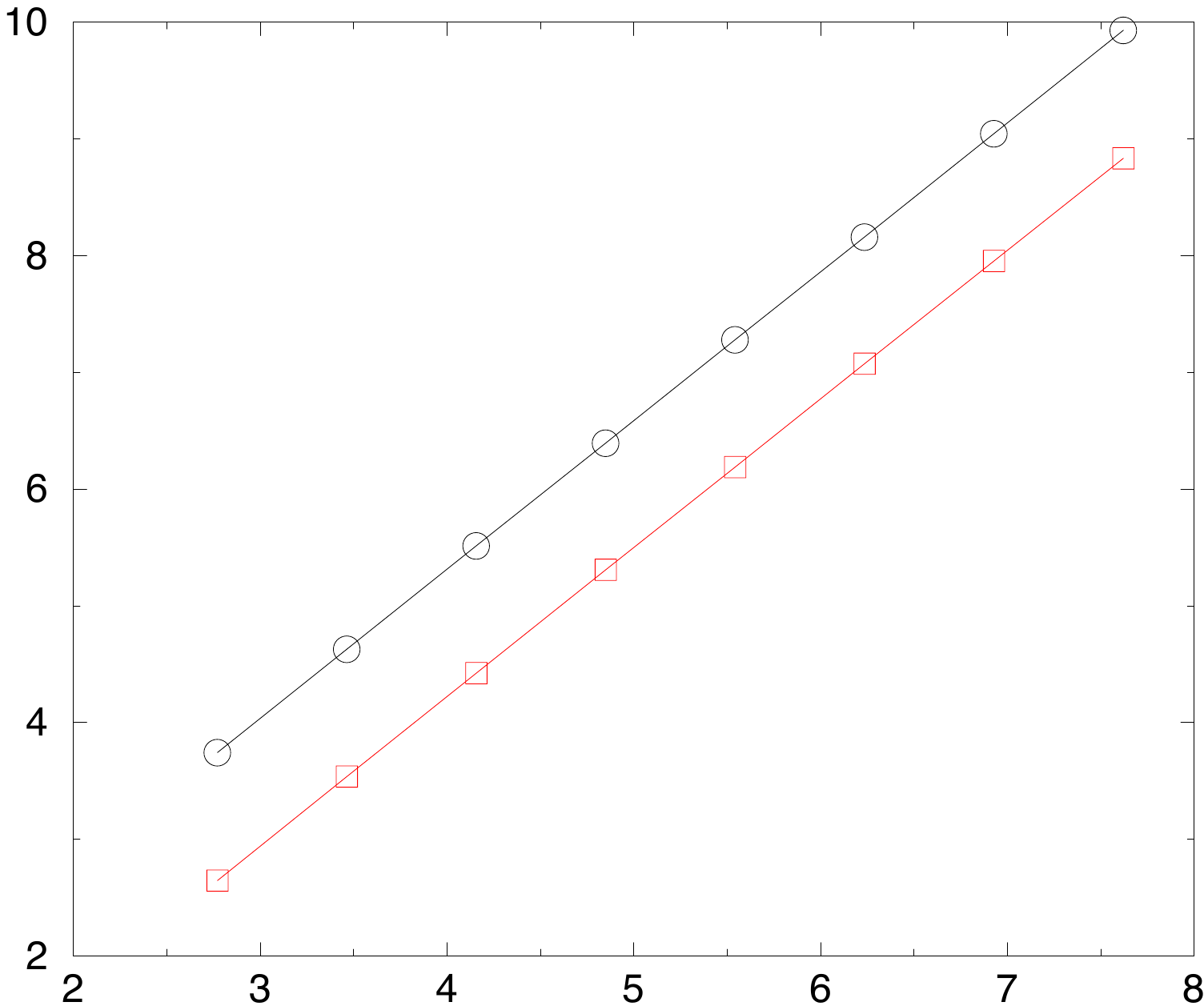}
\caption{ Box-variant of the Strong Disorder Renormalization procedure in two dimensions  : \\
(a) Domain-Wall between the Periodic and the Anti-Periodic Boundary conditions
 in a two dimensional sample of size $2048 \times 2048$. \\
(b) Log-log plot of the average value $ \overline{\Sigma^{DW} } $ (circles)
and of the width $\sqrt{ \overline{ (\Sigma^{DW})^2 }-(\overline{\Sigma^{DW} })^2 } $
(squares) of the length of the Domain-Wall as a function of the size $2^4=16 \leq L \leq 2^{11}=2048$ of samples : the two slopes correspond to the fractal dimension $d_s \simeq 1.27$ as on Fig. \ref{figds2d}.
  }
\label{fig2dbox}
\end{figure}

We have also considered the following Box-variant 
of the Strong Disorder Renormalization procedure.
The initial two-dimensional sample of size $L=2^n$ is first decomposed
into $\left(\frac{L}{2} \right)^2$ boxes of $2^2=4$ spins.
We first eliminate in each box the spin with the highest $\Omega_i$ in the box,
so that there remains three spins per box. 
We then eliminate again in each box the spin with the highest $\Omega_i$ in the box, so that there remains two spins per box. 
We finally eliminate again in each box the spin with the highest $\Omega_i$ in the box, so that there remains one spin per box.
We may now group together four boxes to iterate the procedure.
This variant allows to consider much bigger sizes $L$ and statistics $n_s(L)$
with respect to Eq. \ref{nume2d}
\begin{eqnarray}
 L && = 2^4, 2^5, 2^6, 2^7, 2^8, 2^9, 2^{10}, 2^{11} \nonumber \\
n_s(L) && = 6 \times 10^8,  15 \times 10^7 , 33 \times 10^6 , 75 \times 10^5 ,
2\times 10^6, 4 \times 10^5, 7\times 10^4, 12 \times 10^3
\label{nume2dbox}
\end{eqnarray}
As an example, we show on Fig. \ref{fig2dbox} (a)
 the interface obtained in a given sample of size $2048 \times 2048$.
The statistics over samples of length of the Domain-Wall
corresponds to the same fractal dimension $d_s \simeq 1.27$ as in Eq. \ref{numeds2d}.

The corresponding droplet exponent measured from Eq. \ref{theta2dmes}
is slightly negative 
\begin{eqnarray}
\theta \simeq -0.09
\label{theta2d}
\end{eqnarray}
i.e. it is still far from the correct value $\theta
(d=2) \simeq -0.28$ quoted in Eq. \ref{thetahypercubic}.

\section{ Application to the Gaussian spin-glass in dimension $d=3$ }

\label{sec_rg3d}

\subsection{ Strong Disorder renormalization procedure}

For each disordered sample defined on a cubic lattice $L \times L \times$
\begin{eqnarray}
H_{3d} = -\sum_{x=1}^L \sum_{y=1}^L \sum_{z=1}^L 
S_{(x,y,z)} \left[ J^{P}_x(x,y,z)S_{(x+1,y,z)} + J^{P}_y(x,y,z)S_{(x,y+1,z)}
+ J^{P}_z(x,y,z)S_{(x,y,z+1)}  \right]
\label{hsg3d}
\end{eqnarray}
we have applied the same procedure as in $d=2$ (see details in section \ref{secnum2d}), the Anti-Periodic boundary conditions corresponding again 
 to the change of the signs of the horizontal couplings in the column $x=1$
\begin{eqnarray}
 J_x^{AP}(x=1,y,z) = - J_x^{P}(x=1,y,z)
\label{ruleAP3d}
\end{eqnarray}

\begin{figure}[htbp]
 \includegraphics[height=6cm]{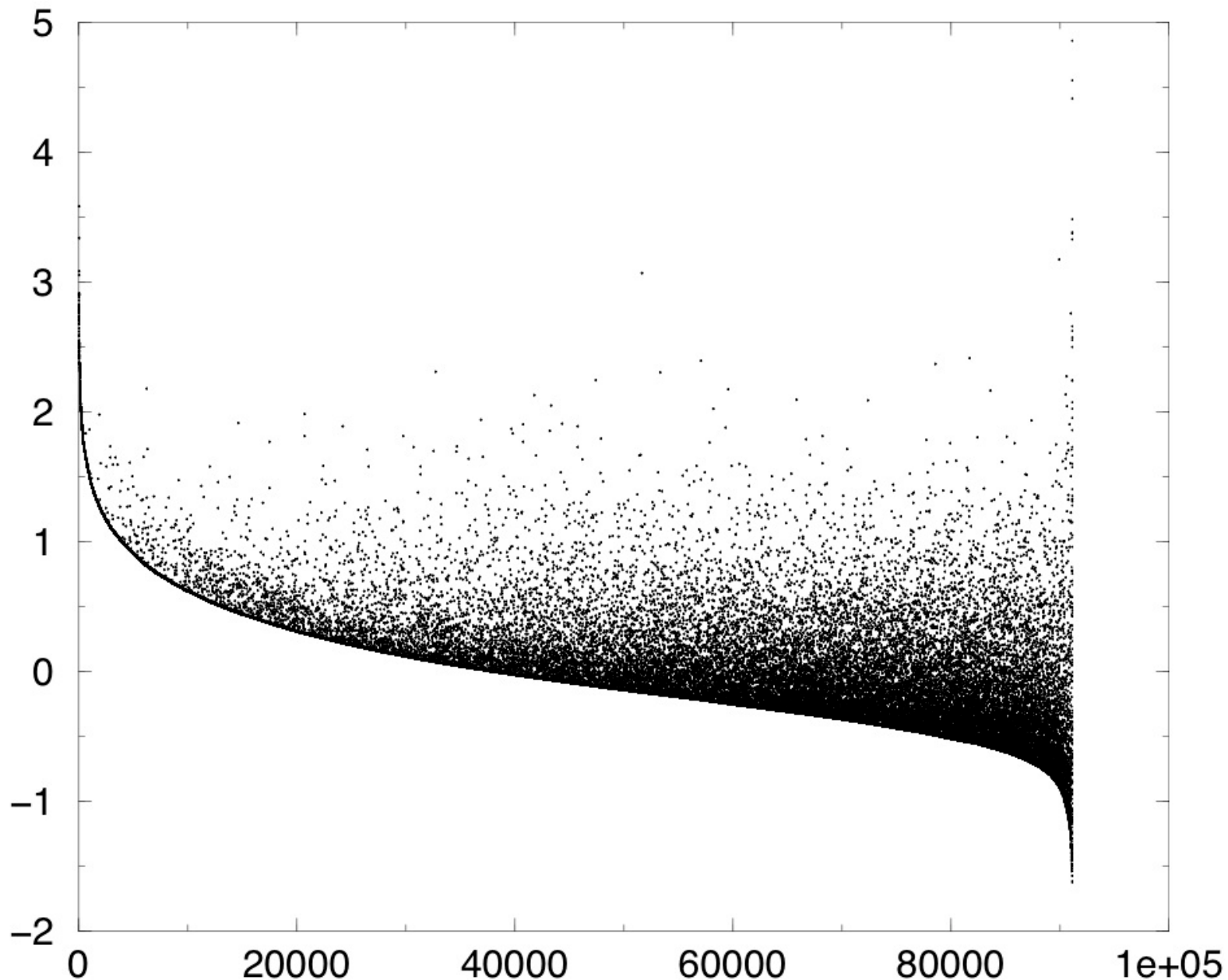}
\hspace{1cm}
 \includegraphics[height=6cm]{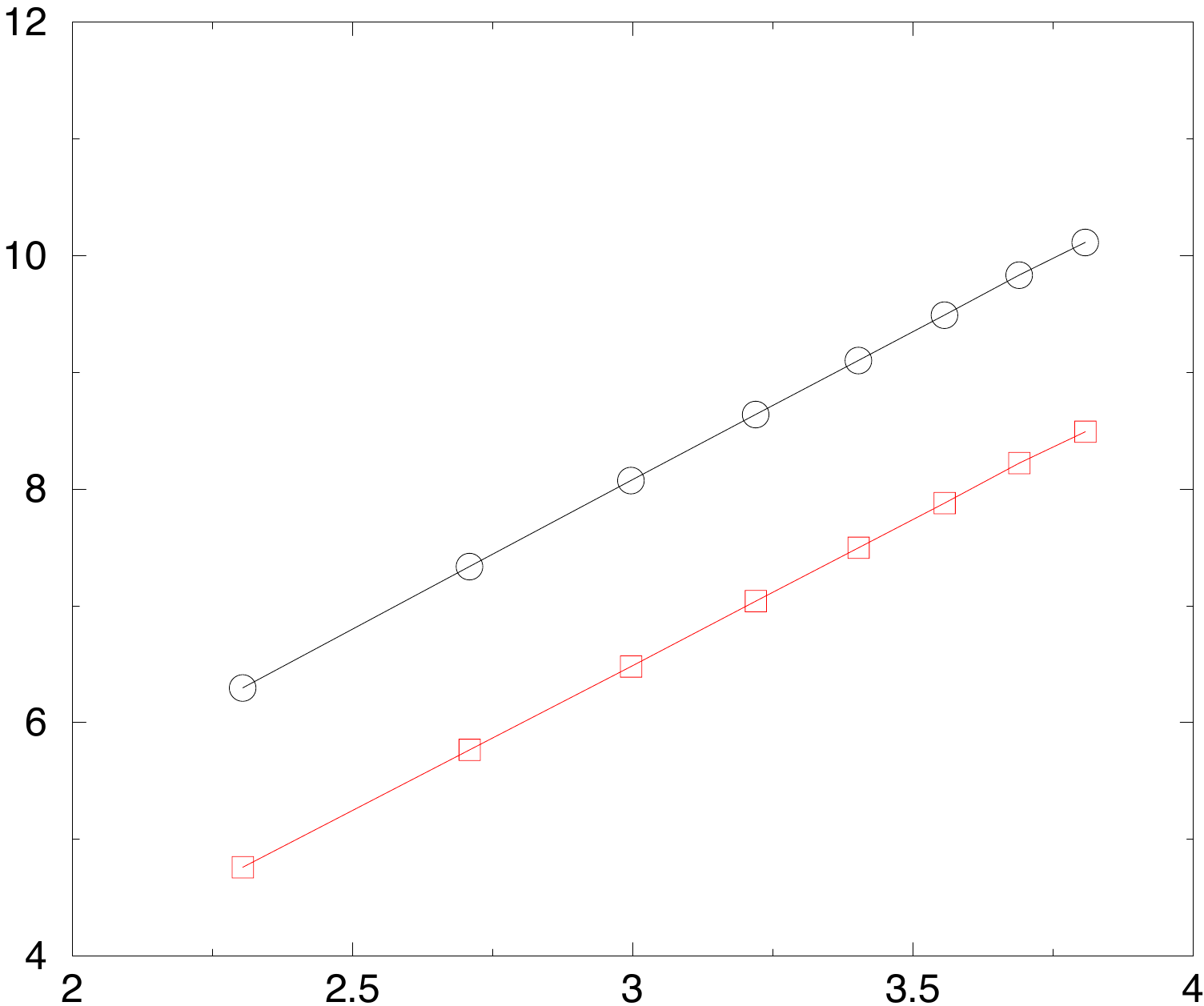}
\caption{ 
Strong Disorder RG procedure in $d=3$ \\
(a) RG parameter $\Omega$ of the decimated spin as a function of the RG step
(the RG step corresponds to 
the number of spins that have already been decimated)
in a sample of size $45 \times 45 \times 45$. \\ 
(b) Log-log plot of the average value $ \overline{\Sigma^{DW} } $ (circles)
and of the width $\sqrt{ \overline{ (\Sigma^{DW})^2 }-(\overline{\Sigma^{DW} })^2 } $
(squares) of the surface of the Domain-Wall as a function of the size $10 \leq L \leq 45$ of samples : the slopes correspond to the fractal dimension $d_s \simeq 2.55$.  }
\label{figds3d}
\end{figure}

The application to $n_s(L)$ independent disordered samples
of various sizes $L$ with
\begin{eqnarray}
 L && = 10, 15, 20, 25, 30, 35, 40, 45 \nonumber \\
n_s(L) && = 21 \times 10^6,  25 \times 10^5 , 38 \times 10^4 ,  10^5 ,
24 \times 10^3, 9 \times 10^3, 3\times 10^3, 16 \times 10^2
\label{nume3d}
\end{eqnarray}
yields that the average value 
and the width of the surface $\Sigma^{DW} $ of the Domain-Wall
have the same scaling (see Fig. \ref{figds3d})
\begin{eqnarray}
 \overline{\Sigma^{DW} } && \propto L^{d_s}  \nonumber \\
\sqrt{ \overline{ (\Sigma^{DW})^2 }-(\overline{\Sigma^{DW} })^2 } && \propto L^{d_s}
\label{sigma3d}
\end{eqnarray}
with the fractal dimension
\begin{eqnarray}
d_s \simeq 2.55
\label{numeds3d}
\end{eqnarray}
in agreement with the value quoted in Eq. \ref{dshypercubic} measured via other numerical methods \cite{palassini,katzgraber}.

\subsection{ Box-variant of the Strong Disorder renormalization procedure }

\begin{figure}[htbp]
 \includegraphics[height=6cm]{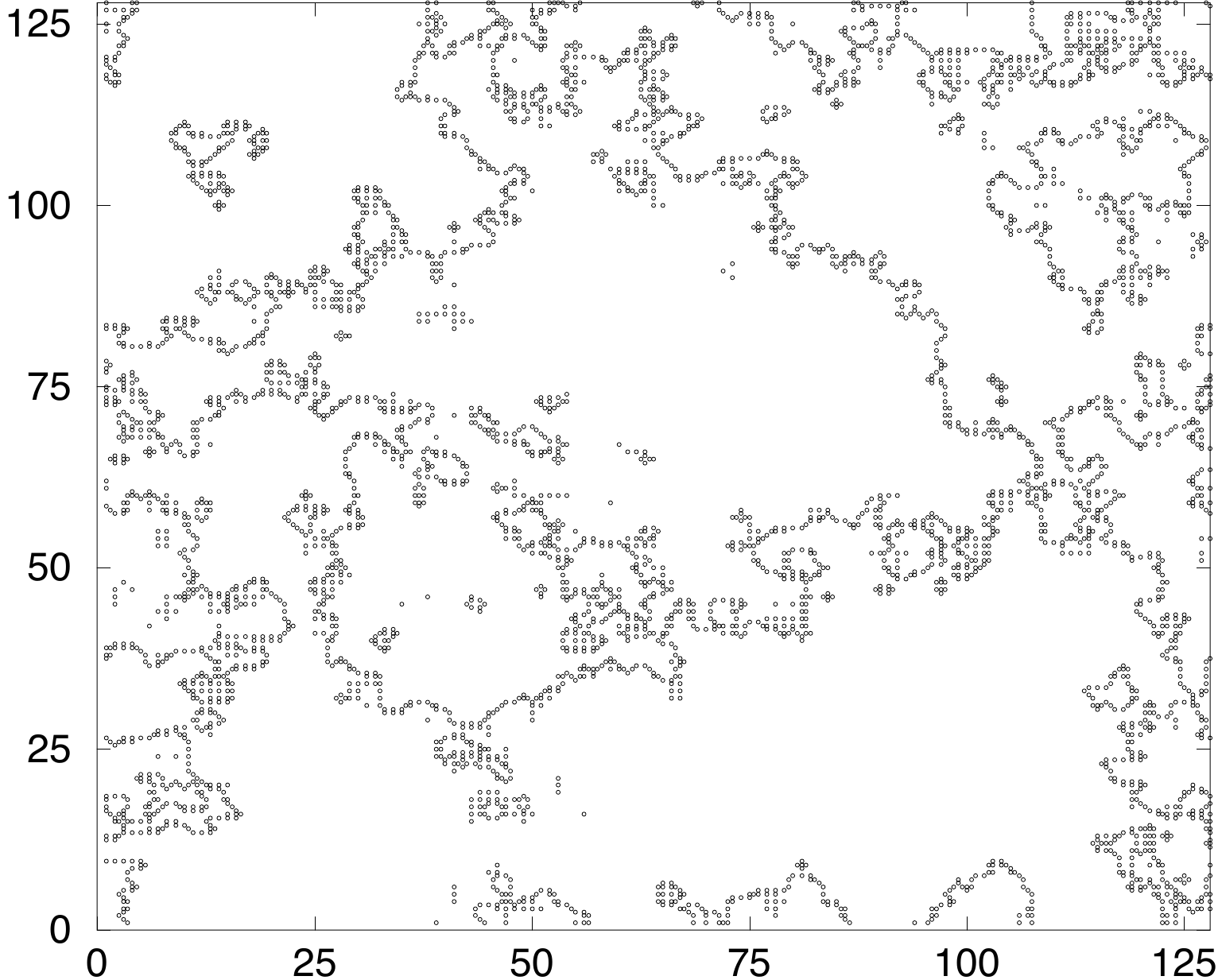}
\hspace{1cm}
 \includegraphics[height=6cm]{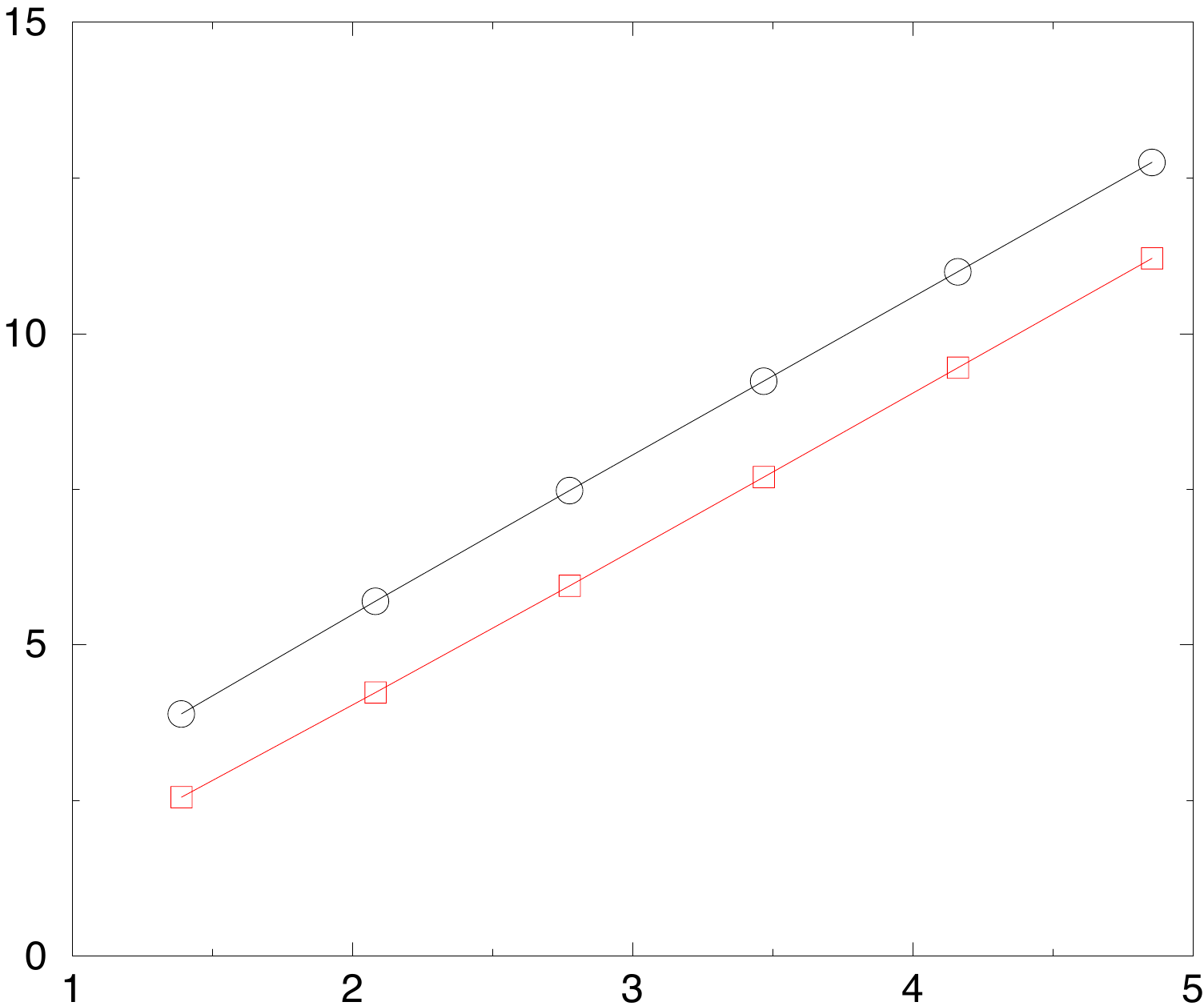}
\caption{ Box-variant of the Strong Disorder Renormalization procedure in three dimensions  : \\
(a) Cut by the plane $x=constant$ containing the maximal number of points
of the Domain-Wall between the Periodic and the Anti-Periodic Boundary conditions
 in a given sample of size $128^3 $.
\\
(b) Log-log plot of the average value $ \overline{\Sigma^{DW} } $ (circles)
and of the width $\sqrt{ \overline{ (\Sigma^{DW})^2 }-(\overline{\Sigma^{DW} })^2 } $
(squares) of the surface of the Domain-Wall as a function of the size
 $2^2=4 \leq L \leq 2^{7}=128$ of samples : the slopes correspond to the same fractal dimension $d_s \simeq 2.55$ as on Fig. \ref{figds3d}.
  }
\label{fig3dbox}
\end{figure}

As in dimension $d=2$ (see section \ref{secboxvariant2d}), 
we have also considered  the following Box-variant 
of the Strong Disorder Renormalization procedure.
The initial three-dimensional sample of linear size $L=2^n$ is first decomposed
into $\left(\frac{L}{2} \right)^3$ boxes of $2^3=8$ spins.
We perform seven sweeps to decimate 
iteratively the spin with the highest $\Omega_i$ in each box,
 so that there remains one spin per box.
We then group together $8$ boxes to iterate the procedure.
This variant allows to consider bigger sizes $L$ and statistics $n_s(L)$
with respect to Eq. \ref{nume3d}
\begin{eqnarray}
 L && = 2^2, 2^3, 2^4, 2^5, 2^6, 2^7 \nonumber \\
n_s(L) && = 15 \times 10^8,  12 \times 10^7 , 13 \times 10^6 ,  10^6 ,
135 \times 10^3, 8  \times 10^3
\label{nume3dbox}
\end{eqnarray}

In contrast to $d=2$ where the Domain-Walls can be easily shown as on Fig.
\ref{figinter2d} (a), we have not found how to represent clearly
the Domain Wall of a given three dimensional sample on a two-dimensional figure.
We have thus chosen to show on Fig. \ref{fig3dbox} (a) the cut of the Domain Wall
by the plane $x=constant$ where the number of points is maximum.
As shown on Fig. \ref{fig3dbox} (b),
the statistics over samples of the surface of the Domain-Wall
corresponds to the same fractal dimension $d_s \simeq 2.55$ as in Eq. \ref{numeds2d}.

The droplet exponent that we measure from the width of the distribution
of the Domain-Wall energy $ E^{DW}$ (Eq. \ref{theta2dmes})
\begin{eqnarray}
\theta \simeq 0.75
\label{theta3d}
\end{eqnarray}
 is again far from the correct value $\theta
(d=3) \simeq 0.24$ quoted in Eq. \ref{thetahypercubic}, even if it is not as bad
as the Block value $\theta^{Block}(d=3) = 1 $ of Eq. \ref{thetablock}.

\section{ Conclusion  }

\label{sec_conclusion}

In summary, we have introduced and studied numerically a
 simple Strong Disorder renormalization procedure at zero temperature for spin-glasses in dimension $d=2$ and $d=3$.
Our main conclusion is that it is able to reproduce very
 well the fractal dimensions $d_s$ quoted in Eq. \ref{dshypercubic},
although it is not able to reproduce the correct droplet exponents of Eq. \ref{thetahypercubic}. A possible interpretation is that the fractal dimension $d_s$
is actually determined by the short-scales
 optimization well captured by the simple 
Strong Disorder RG procedure, whereas the droplet exponent $\theta$ is determined
by large-scales optimization that is not well captured by the simple 
Strong Disorder RG procedure, because the RG parameter $\Omega$ remains positive during the first part of the RG steps but tends to become negative in the last part  (see Figs \ref{figinter2d} (b) and \ref{figds3d}).
This situation is thus opposite to the Migdal-Kadanoff renormalization,
which reproduces very well the droplet exponent $\theta$ in dimensions $d=2$ and $d=3$,
but not the surface fractal dimension $d_s$ (Eq. \ref{dsMK}).
Let us hope that in the future it will be possible to formulate an RG procedure able to reproduce both exponents $(\theta,d_s)$ correctly!

\end{document}